\documentclass[referee]{aa} 
\usepackage{graphicx} 
\begin{document}
   \title{Relativistic fine structure oscillator strengths for Li-like ions: 
C IV - Si XII, S XIV, Ar XVI, Ca XVIII, Ti XX, Cr XXII, and Ni XXVI}
  

   \author{Sultana N. Nahar
          \thanks{Complete electronic files for energies (e.g. Tables 3a 
and 3b) and transition probabilities (e.g. Table 7) are available at the CDS 
via anonymous ftp to cdsarc.u-strasbg.fr (130.79.128.5) }
          }

   \offprints{S.N. Nahar}

   \institute{Department of Astronomy, The Ohio State University,
              Columbus, OH 43210, USA \\
              \email{nahar@astronomy.ohio-state.edu}
             }


   \abstract{
Ab initio calculations including relativistic effects employing
the Breit-Pauli R-matrix (BPRM) method are reported for fine structure
energy levels and oscillator strengths upto $n$ = 10 and 0 $\leq l 
\leq$ 9 for 15 Li-like ions: C IV, N V, O VI, F VII, Ne VIII, Na IX, 
Mg X, Al XI, Si XII, S XIV, Ar XVI, Ca XIII, Ti XX, Cr XXII, and Ni 
XXVI. About one hundred bound fine structure energy levels of total 
angular momenta, 1/2 $\leq J \leq $ 17/2 of even and odd parities, total 
orbital angular momentum, 0 $\leq L \leq$ 9 and spin multiplicity (2S+1) 
= 2, 4 are considered for each ion. The levels provide almost 900 dipole 
allowed and intercombination bound-bound transitions. The BPRM method 
enables consideration of large set of transitions with uniform accuracy 
compared to the best available theoretical methods. The CC eigenfunction 
expansion for each ion includes the lowest 17 fine structure energy 
levels of the core configurations $1s^2$, $1s2s$, $1s2p$, $1s3s$, $1s3p$, 
and $1s3d$. The calculated energies of the ions agree with the measured 
values to within 1\% for most levels. The transition probabilities show 
good agreement with the best available calculated values. The results 
provide the largest sets of energy levels and transition rates for the 
ions and are expected to be useful in the analysis of X-ray and EUV 
spectra from astrophysical sources.

   }

%

\section{Introduction}

A wealth of high resolution astrophysical spectra are being obtained 
by ground-based telescopes and by space based observatories such as HST, 
CHANDRA, ISO, FUSE. Accurate spectral analysis provides diagnostic 
of element abundances, temperatures etc.. However, a major task is the 
identification of the large number of lines, especially from UV to X-ray 
region for use in sythetic models, calculating opacities. Ab initio 
relativistic calculations using the Breit-Pauli R-matrix (BPRM) method, 
developed under the Iron Project (IP, Hummer et al. 1993), are carried 
out for extensive and accurate sets of oscillator strengths ($f$), line 
strengths ($S$)and radiative transition probabilities ($A$) for a number 
of Li-like ions from carbon to nickel. Results for lithium like Fe XXIV
were reported earlier (Nahar and Pradhan 1999). Compared to the very 
accurate theoretical methods for oscillator strengths for a relatively 
small number of transitions, the BPRM method allows consideration of 
a large number of transitions with comparable accuracy for most of the 
transitions. 

Relatively smaller sets of transitions are available for the lithium 
like ions considered. An evaluated compilation of the results by 
various investigators obtained using various approximation is available
from the web based database of the National Institute for Standards and 
Technology (NIST). The previous large sets of non-relativistic data were 
obtained by Peach et al. (1988) under the Opacity Project (OP 1995, 1996) 
which are accessible through the OP database, TOPbase (Cunto et al. 1993). 
Nahar (1998) obtained later a larger set of transitions for O VI using 
a larger wavefunction expansion. These results consider only the dipole 
allowed $LS$ multiplets, i.e., no relativistic fine structure 
splitting were taken into account. The OP datasets for a number of ions 
have been reprocessed to obtain fine structure oscillator strengths 
through pure algebraic transformation of the line strengths and 
utilizing the observed energies for improved accuracy, such as, the 
recent compilation of transition probabilities by NIST for C, N, O ions 
(Wiese et al 1996), the transition probabilities for Fe~II by Nahar (1995).

\section{Theory }

Theoretical details are discussed in previous works, such as in the 
first large scale relativistic calculations using the BPRM method for 
bound-bound transitions in Fe XXIV and Fe XXV (Nahar and Pradhan 1999). 
The close coupling (CC) approximation using the R-matrix method as 
employed under the OP (Seaton 1987, Berrington et al. 1987) was extended 
to BPRM method under the Iron Project (IP, Hummer et al. 1993) to
include the relativistic effects in the Breit-Pauli approximation 
(Scott and Burke 1980, Scott and Taylor 1982, Berrington et al. 1995).
They are derived from atomic collision theory using the coupled channel 
approximation. The BPRM method has been used for several other ions, 
such as Fe V (Nahar et al. 2000), Ar XIII and Fe XXI (Nahar 2000), C II 
and C III (Nahar 2002), C III (Berrington et al. 2001), Na III 
(Berrington 2001), and Cl-like ions (Berrington et al. 2001).

In the CC approximation the wavefunction expansion, $\Psi(E)$, for a 
(N+1) electron system with total spin and orbital angular momenta
symmetry  $SL\pi$ or total angular momentun symmetry $J\pi$, is 
described in terms of the target ion states as:

\begin{equation}
\Psi_E(e+ion) = A \sum_i \chi_i(ion)\theta_i + \sum_j c_j \Phi_j(e+ion),
\end{equation}

\noindent
where $\chi_{i}$ is the target ion wavefunction in a specific state
$S_iL_i\pi_i$ or level $J_i\pi_i$, and $\theta_{i}$ is the wavefunction 
for the interacting (N+1)th electron in a channel labeled as 
$S_iL_i(J_i)\pi_i \ k_{i}^{2}\ell_i(SL\pi~or~ \ J\pi)$; $k_{i}^{2}$ is the
incident kinetic energy. In the second sum the $\Phi_j$'s are correlation
wavefunctions of the (N+1) electron system that (a) compensate for the
orthogonality conditions between the continuum and the bound orbitals,
and (b) represent additional short-range correlation that is often of
crucial importance in scattering and radiative CC calculations for each
$SL\pi$. 

The relativistic (N+1)-electron Hamiltonian for the N-electron target
ion and a free electron in the Breit-Pauli approximation, as adopted under
the IP, is 
\begin{equation}
H_{N+1}^{\rm BP}=H_{N+1}+H_{N+1}^{\rm mass} + H_{N+1}^{\rm Dar}
+ H_{N+1}^{\rm so},
\end{equation}
where $H_{N+1}$ is the non-relativistic Hamiltonian,
\begin{equation}
H_{N+1} = \sum_{i=1}\sp{N+1}\left\{-\nabla_i\sp 2 - \frac{2Z}{r_i}
        + \sum_{j>i}\sp{N+1} \frac{2}{r_{ij}}\right\} . 
\end{equation}
added by the one-body mass correction term, the Darwin term and the 
spin-orbit interaction term. The mass-correction and Darwin terms do not 
break the LS symmetry, while the spin-orbit interaction split the LS 
terms into fine-structure levels labeled by $J\pi$. The BP Hamiltonian in 
the present work does not include the full Breit-interaction in that the 
two-body spin-spin and spin-other-orbit terms are not included. 

The set of ${SL\pi}$ are recoupled to obtain (e + ion) states  with 
total $J\pi$, following the diagonalization of the (N+1)-electron 
Hamiltonian, 
\begin{equation}
H^{BP}_{N+1}\mit\Psi = E\mit\Psi.
\end{equation}
Substitution of the wavefunction expansion introduces set of coupled
equations that are solved using the R-matrix approach. The details of the 
solutions for the wavefunctions and energies can be found in the OP 
papers (1995) and in Hummer et al. (1993). The channels, characterized by 
the spin and angular quantum numbers of the (e + ion) system, describe 
the scattering process with the free electron interacting with the 
target at positive energies (E $>$ 0), while at {\it negative} total 
energies (E $<$ 0), the solutions of the close coupling equations occur 
at discrete eigenvalues of the (e + ion) Hamiltonian that correspond to 
pure bound states $\Psi_B$.

The oscillator strength ($f$-values) for a bound-bound transition can be 
obtained from the transition matrix,
\begin{equation}
<\Psi_B || {\bf D} || \Psi_{B'}>,
\end{equation}
where ${\bf D}$ is the dipole operator. The transition matrix can be 
reduced to the generalised line strength ($S$), in either length or 
velocity form as
\begin{equation}
S_{\rm L}=
 \left|\left\langle{\mit\Psi}_f
 \vert\sum_{j=1}^{N+1} r_j\vert
 {\mit\Psi}_i\right\rangle\right|^2 \label{eq:SLe}
\end{equation}
and
\begin{equation}
S_{\rm V}=\omega^{-2}
 \left|\left\langle{\mit\Psi}_f
 \vert\sum_{j=1}^{N+1} \frac{\partial}{\partial r_j}\vert
 {\mit\Psi}_i\right\rangle\right|^2, \label{eq:SVe}
\end{equation}
where $\omega$ is the incident photon energy in Rydberg units, 
$\mit\Psi_i$ and $\mit\Psi_f$ are the initial and final state bound
wavefunctions respectively. 

In terms of the transition energy $E_{ji}$ between states i and j,
the oscillator strength, $f_{ij}$, is obtained from $S$ as

\begin{equation}
f_{ij} = {E_{ji}\over {3g_i}}S,
\end{equation}

\noindent
and the transition probability or Einstein's A-coefficient, $A_{ji}$, as

\begin{equation}
A_{ji}(a.u.) = {1\over 2}\alpha^3{g_i\over g_j}E_{ji}^2f_{ij},
\end{equation}

\noindent
where $\alpha$ is the fine structure constant, and $g_i$, $g_j$ are the
statistical weight factors of the initial and final states, respectively. 
The lifetime of a level can be obtained from the A-values of the level as,
\begin{equation}
\tau_k(s) = {1\over A_k},
\end{equation}
where $A_k$ is the total radiative transition probability for the
level k, i.e., $A_k = {\sum_i A_{ki}(s^{-1})}$, and $A_{ji}(s^{-1}) = 
{A_{ji}(a.u.)/\tau_0}$, $\tau_0 = 2.4191\times 10^{-17}$s is the 
atomic unit of time.

\section{Atomic calculations}

The Breit-Pauli R-matrix calculations for the Li-like ions are 
carried out using an eigenfunction expansion of 17 fine 
structure levels of configurations, $1s^2$, $1s2s$, $1s2p$, $1s3s$, 
$1s3p$ and $1s3d$ of the He-like target or core (Table 1) for each ion. 

The orbital wavefunctions of the target are obtained from the atomic 
structure calculations using the code SUPERSTRUCTURE (Eissner et al. 
1974) that employs Thomas-Fermi potential. The wavefunctions of the 
spectroscopic levels are optimized individually for each ion. $4s$ and 
$4p$ are treated as correlation orbitals. The optimization is carried 
out such that the set of configurations and Thomal Fermi scaling 
parameters ($\lambda$) for the orbitals yield calculated level energies 
that agree closely with the measured values and the discrepancy between 
the length and velocity form oscillator strengths is less than 5\% for 
the allowed transitions from the ground level. While the set of 
spectroscopic configurations remains the same for each ion, the set of 
correlation configurations and parameters $\lambda$ for the orbitals 
vary for some, as seen in Table 1. The level energies given in the 
table are mainly from the measured values in the database of the NIST. 
The calculated fine structure energies differ by much less than 1\% from 
the  measured values of the levels. However, in the R-matrix 
calculations the calculated energies are replaced by the observed 
values whenever available, i.e., calculated energies are used only 
when the measured values are not available. 

For the (N+1)th electron, all partial waves of 0$\leq l\leq $ 9 are 
included. The bound-channel term of the wavefunction, the second term 
in Eq. (1), includes all possible (N+1)-configurations from a 
vacant shell to maximum occupancies of $1s^2$, $2s^2$, $2p^2$, $3s^2$, 
$3p^2$, $3d^2$, $4s$, and $4p$. 

The BPRM calculations consist of several stages of computation 
(Berrington et al. 1995). The orbital wavefuntions of SUPERSTRUCTURE are 
used as the input for the BPRM codes to compute the one- and two-electron 
radial integrals. The R-matrix basis set consists of 30 continuum functions 
for each ion. The calculations included all possible bound levels for 
0.5 $\leq J\leq$ 8.5 of even and odd parities, with $n < 10, \ \ell 
\leq 9$, $0\leq L \leq$ 11 or 12, and $(2S+1)$=2, 4. The intermediate 
coupling calculations are carried out on recoupling the $LS$ symmetries 
in a pair-coupling representation in stage RECUPD. The (e + core) 
Hamiltonian matrix is diagonalized for each resulting $J\pi$ in STGH.

The fine structure bound levels are sorted through the poles in the 
(e + ion) Hamiltonian with a fine mesh of effective quantum number 
$\nu$. The mesh ($\Delta \nu$ = 0.001) is finer than that typically
used for $LS$ energy terms ($\Delta \nu$ = 0.01) to avoid any missing 
levels and to obtain accurate energies for the higher levels. 

About a hundred fine structure bound energy levels are obtained for each 
ion. They are obtained as sets of levels belonging to symmetries $J\pi$ 
only, complete spectroscopic designations for identifications are not 
specified. The level identification scheme, based on quantum defect 
analysis and percentage of channel contributions to the levels, as 
developed in the code PRCBPID (Nahar and Pradhan 2000) is employed. 
Hund's rule is used for positions of the levels such that a level with 
higher angular orbital momentum $L$ may lie below the low $L$ one. 
Although level identification of Li-like ions is straight forward, it is 
more nvolved for complex ions. The final designation is given by 
$C_t(S_tL_t\pi_t)J_tnlJ(SL)\pi$ where $C_t$, $S_tL_t\pi_t$, $J_t$ are 
the configuration, $LS$ term and parity, and total angular momentum of 
the target, $nl$ are the principal and orbital quantum numbers of the 
outer or the valence electron, and $J$ and $SL\pi$ are the total angular 
momentum, $LS$ term and parity of the $(N+1)$-electron system.

\section{Results and Discussions}

Extensive sets of fine structure energy levels and oscillator strengths 
and transition probabilities for the bound-bound transitions are obtained 
for 15 Li-like ions: C IV, N V, O VI, F VII, Ne VIII, Na IX, Mg X, Al XI, 
Si XII, S XIV, Ar XVI, Ca XIII, Ti XX, Cr XXII, and Ni XXVI. The energy 
levels and bound-bound transitions are discussed separately in the two
following sections.

\subsection{Fine Structure Energy Levels}

A total of about 98 fine structure energy levels are obtained for each 
15 Li-like ion (97 or 99 for a few of them). They correspond to levels 
of 2 $\leq n\leq $ 10 and 0 $\leq l \leq $ 9 with total angular momentum, 
1/2 $\leq J \leq $ 17/2 of even and odd parities, total spin multiplicity 
2S+1 = 2, and total orbital angular momentum, 0 $\leq L \leq$ 9. 
All levels have been identified. The number of levels obtained far exceed 
the observed or previously calculated ones.

The calculated energies are compared in Table 2 with the measured 
values, compiled by the NIST. The table presents comparison of energies 
of a few ions, such as C IV, O VI, and Ni XXVI, as examples. 
The calculated energies of each ion agree very well with the measured 
values, within 1\% for almost all levels, and for all the ions. 
For levels with $L \geq$ 4 there is nearly exact agreement, as 
expected for hydrogenic behavior of the highly excited states.
These are the most detailed close coupling calculations for these ions. 
The complete energy levels of the 15 ions are availabe eletronically.

The complete set of energies are presented in two formats, as in the case 
for other ions obtained previously, e.g. for Fe V (Nahar et al.  2000), 
for consistency.  One is in LS term format where the fine structure 
components of a $LS$ term are grouped together, useful for spectroscopic 
diagnostics. Table 3a presents sample of the table containing total 
sets of energies. The table contains partial set of levels 
of C IV and Ni XXVI. For each set of levels, the columns provide the 
core information, $C_t(SL\pi~J)_t$, the configuration of the outer 
electron, $nl$, total angular momentun, $J$, energy in Rydberg, the 
effective quantum number of the valence electron, $\nu$, and the $LS$ 
term designation of the level. The top line of the set gives the number 
of fine structure levels expected ($Nlv$), followed by the spin and 
parity of the set ($^{2S+1}L^{\pi}$), followed by the values of $L$, 
where values of the total angular momentum $J$, associated with each 
$L$, are given within parentheses. The last line gives the number of 
calculated levels ($Nlv(c)$) obtained with a statement of completeness 
of the calculated set.

In the other format, the fine structure levels are presented in sets 
belonging to different $J\pi$ symmetries where levels are in energy 
order as shown in sample table, Table 3b. The format is convenient for 
easy implementation in astrophysical models requiring large number of
energy levels and the corresponding transitions. At the top of each 
set. the total number of energy levels ($Nlv$) and the symmetry 
information $J\pi$ are given. For example, there are 9 fine structure 
levels of C IV with $J\pi$ = $0^e$. The levels are identified with the 
configuration and $LS$ term of the core, the outer electron quantum 
numbers, energy, the effective quantum number ($\nu$), and the $LS$
term designation. $\nu~=~z/\sqrt(E-E_t)$ where $E_t$ is the next
immediate target threshold energy.  

Table 4 lists the energies of the eight fine structure levels of n = 2 
and 3 complexes of all ions from C IV to Ni XXVI. These levels are of
astrophysical interest as they are often displayed in the spectra. 
They have been singled out to present the oscillator strengths for 
transitions among them.

\subsection{Oscillator strengths}

About nine hundred oscillator strengths are obtained for the allowed 
transitions in each Li-like ion. Astrophysical models, such as for 
stellar opacity calculations, will require all possible transitions for 
$n$ going upto 10. However, spectral diagnostics may involve only the 
lowest transitions. 

Table 5 presents the oscillator strengths (f) and the transition
probabilities ($A$) for transitions among $n$ = 2 and 3 levels for each
15 ions from C IV to Ni XXVI. There are 14 such transitions for the 8 fine
structure levels as presented in Table 4. Here the energies are expressed
in transition wavelengths rather than individual level energies in 
Rydberg since wavelengths are often used in astrophysical spectral
analysis. However, these transition wavelengths are calculated from 
the measured level energies, as given in the NIST compiled table,
and using the conversion factor, 1 Ry = 911.2671 $\AA$. Hence, the $f$ 
and $A$-values in this table are slightly different from those in the 
original calculated set where calculated transition energies are used. 
The energy independent line strengths $S$ remain the same in both sets. 
Since the difference between the calculated and measured energies is 
typically less than 1\%, the reprocessed set in Table 5 has slight 
improved accuracy. 
 
The BPRM $A$-values are compared with the available data, mainly from 
the compiled table by the NIST where they rate the accuracy to be less 
than 10\%. The lifetimes of some levels of Li-like ions have
been measured and can be obtained theoretically from the sum of the $A$
values as mentioned above. Lifetime experiments have been carried out by,
for example, Heckmann et al.  (1976), Pinnington et al. (1974) for O VI, 
Tra\~{a}bert et al. (1977) for Si XII using beam-foil technique. However, 
present comparison is made mainly with individual $A$-values. Table 6 
shows that present $A$-values are in very good agreement with the highly 
rated compiled values by NIST indicating that present $A$-values can be 
estimated to be accurate at least within 10\%.

Transition probabilities ($A$) for a few ions, such as S XIV, Ar XVI, 
and Ca XVII are not available in the NIST compilation. $A$-values 
for these ion are compared with those obtained from very accurate 
relativistic third-order many-body perturbation theory by Johnson et 
al. (1996). Present values agree almost exactly with those by Johnson 
et al (1996) for the two transitions $2s(^2S_{1/2})-2p(^2P^o_{1/2,3/2})$, 
and by less than 5\% for the two transitions, 
$2p(^2P^o_{1/2,3/2})-3s(^2S_{1/2})$.  Yan et al (1998) have calculated 
the level energies and oscillator strengths for lithium like ions up to 
Z = 20 using Hylleras type variational method including finite nuclear 
mass effects. They present nonrelativistic $1s^22s(^2S)-1s^22p(^2P^o)$ 
oscillator strengths for the ions. Present weighted averaged $f$-values 
for S XIV, Ar XVI, and Ca XXVII are compared with those by Yan et al. 
at the end of Table 6. Present f-values agree within 10\% with the 
nonrelativistic values by Yan et al.

The agreement between the present values and those from previous 
calculations indicates that the higher order relativistic and QED terms 
omitted from the BP Hamiltonian (Eq. 2) may not affect the transition 
probabilities of the ions considered herein by more than a few percent. 

The complete set of fine structure transitions for the ions are available 
electronically. The tables contain calculated transition probabilities 
($A$), oscillator strengths ($f$), and line strengths ($S$). The 
calculated level energies are also given in the same table. A sample set 
of transitions 
is presented in Table 7 for O VI. The top of the table specifies the 
nuclear charge (Z = 8) and number of electrons in the ion, $N_{elc}$ 
(= 3). Below this line are the sets of oscillator strengths belonging 
to pairs of symmetries, $J_i\pi_i~-~J_k\pi_k$. The symmetries are 
expressed in the form of $2J_i$ and $\pi_i$ ($\pi$=0 for even and =1 
for odd parity), $2J_k$ and $\pi_k$, at the top of the set. 
For example, Table 7 present partial transitions for two pairs of 
symmetries, $J=1/2^e-J=1/2^o$ and $J=1/2^e-J=3/2^o$ of O VI. The line 
below the symmetries gives the number of bound levels of the two
transitional symmetries, $N_{Ji}$ and $N_{Jk}$. The line is followed 
by $N_{Ji}\times N_{Jk}$ number of transitions. The first two columns 
are the energy level indices, $I_i$ and $I_k$, whose identification 
can be found from the energy table, the third and the fourth 
columns provide the energies, $E_i$ and $E_k$, in Rydberg unit. 
The fifth column is the $gf_L$ for the allowed transitions 
($\Delta J$ = 0,$\pm 1$). where $f_L$ is the oscillator strength
in length form, and $g~=~2J+1$ is the statistical weight factor of
the initial or the lower level. A negative value for $gf$ means that
$i$ is the lower level, while a positive one means that $k$ is the
lower level. Column six is the line strength (S) and the last column
is the transition probability, $A_{ki}(sec^{-1})$.

\section{Conclusion}

Accurate and large-scale calculations have been carried out for the
set of fine structure energy levels and transition probabilites upto 
$n$ = 10 for 15 Li-like ions from C IV to Ni XXVI. The set of results 
far exceeds the currently available experimental and theoretical data. 

The results are obtained in intermediate coupling including 
relativistic effects using the Breit Pauli R-matrix method (BPRM) in 
the close coupling approximation. Both the energies and the transition
probabilities show very good agreement, within 1-10\%, with almost all 
accurate calculated and measured values available. This indicates that 
for these highly charged ions the higher order relativistic and QED 
effects omitted in the BPRM calculations may lead to an error not 
exceeding the estimated uncertainty. 

The results from the present work should be particularly useful in 
the analysis of X-ray and Extreme Ultraviolet spectra from 
astrophysical and laboratory sources where non-local thermodynamic 
equilibrium (NLTE) atomic models with many excited levels are needed.

%

\begin{acknowledgements}
This work was partially supported by U.S. National Science Foundation 
(AST-9870089) and the NASA ADP program. The computational work was carried 
out on the Cray T94 and Cray SV1 at the Ohio Supercomputer Center in 
Columbus, Ohio.

\end{acknowledgements}


\begin{table*}
\caption{Fine structure energy levels of the He-like core included in the
eigenfunction expansion of the Li-like ions. The spectroscopic
set of configurations ($1s^2$, $1s2s$, $1s2p$, $1s3s$, $1s3p$, $1s3d$) is
common to each ion. The common correlation configurations are: $2s^2$, 
$2p^2$, $3s^2$, $3p^2$, $3d^2$, $2s2p$, $2s3s$, $2s3p$, $2s3d$, $2s4s$, 
$2s4p$, $2p3s$, $2p3p$, $2p4s$, $2p4p$ while the additional
ones are listed below the table alongwith the values of the Thomas-Fermi 
scaling parameter ($\lambda$) for the orbitals.
}
\begin{tabular}{lrrrrrrrc}
 \noalign{\smallskip}
 \hline
 \noalign{\smallskip}
 Level & \multicolumn{8}{c}{Energy (Ry)} \\
& Z=  6 &   7 &   8 &   9 &  10 &  11 &  12 &  13 \\
 \noalign{\smallskip}
 \hline
 \noalign{\smallskip}
 $1s2       (^1S^e_0)$ &    0.0000 &    0.0000 &    0.0000 &    0.0000 &    0.0000 &    0.0000 &    0.0000 &    0.0000 \\
 $1s2s      (^3S^e_1)$ &   21.9730 &   30.8550 &   41.2320 &   53.1272 &   66.5220 &   81.3907 &   97.8349 &  115.7588 \\
 $1s2p      (^3P^o_0)$ &   22.3731 &   31.3320 &   41.7870 &   53.7702 &   67.2350 &   82.1766 &   98.7083 &  116.7137 \\
 $1s2p      (^3P^o_1)$ &   22.3730 &   31.3320 &   41.7880 &   53.7716 &   67.2380 &   82.1896 &   98.7159 &  116.7250 \\
 $1s2p      (^3P^o_2)$ &   22.3742 &   31.3350 &   41.7930 &   53.7803 &   67.2520 &   82.2172 &   98.7484 &  116.7717 \\
 $1s2s      (^1S^e_0)$ &   22.3718 &   31.3410 &   41.8120 &   53.7930 &   67.2760 &   82.2395 &   98.7702 &  116.7858 \\
\hline
\end{tabular}
\end{table*}

\begin{table}
\noindent{Table 2. Comparison of calculated fine structure energies, 
$E_c$, with the observed values, $E_o$ (NIST) for C IV, O VI and Ni XXVI. \\
}
\begin{tabular}{llrrr}
\hline
\noalign{\smallskip}
\multicolumn{2}{c}{Level} & $J$ & \multicolumn{1}{c}{$E_o$(Ry)} & 
\multicolumn{1}{c}{$E_c$(Ry)} \\
\hline
            \noalign{\smallskip}
\multicolumn{5}{c}{C IV} \\
            \noalign{\smallskip}
            \hline
           \noalign{\smallskip}
 $ 1s22s      $ & $^2S^e$ &   0.5  &  4.74020 &  4.73899  \\
 $ 1s22p      $ & $^2P^o$ &   1.5  &  4.15160 &  4.14885  \\
 $ 1s22p      $ & $^2P^o$ &   0.5  &  4.15260 &  4.15050  \\
 $ 1s23s      $ & $^2S^e$ &   0.5  &  1.98050 &  1.98187  \\
 $ 1s23p      $ & $^2P^o$ &   1.5  &  1.82340 &  1.82478  \\
 $ 1s23p      $ & $^2P^o$ &   0.5  &  1.82370 &  1.82526  \\
\hline
 \end{tabular}
 \end{table}

\begin{table}
\noindent{Table 3a. Sample table of fine structure energy levels of Li-like 
ions as sets of $LS$ term components. $C_t$ is the core configuration, $\nu$ 
is the effective quantum number. \\
}
\scriptsize
\begin{tabular}{lcrrrcrl}
\noalign{\smallskip}
\hline
\noalign{\smallskip}
 \multicolumn{2}{c}{$C_t(S_tL_t\pi_t)$} & $J_t$ & $nl$ & $J$ & E(Ry) & $\nu$ & $SL\pi$ \\
\noalign{\smallskip}
 \hline
\noalign{\smallskip}
\multicolumn{8}{c}{C VI} \\
 \noalign{\smallskip}
 \hline
 \noalign{\smallskip}
 \multicolumn{8}{l}{Nlv=~ 1,~~$^2L^e$:
 S ( 1 )/2                                                               } \\
 \noalign{\smallskip}
 1s2      &(1Se)& 0  & 2s&  1& -4.73899E+00 &  1.84 &  2~ S e       \\
 \noalign{\smallskip}
 \multicolumn{8}{l}{Nlv(c)=~ 1~ : set complete                                                                                                         } \\
 \noalign{\smallskip}
 \hline
 \noalign{\smallskip}
 \multicolumn{8}{l}{Nlv=~ 2,~~$^2L^o$:
 P ( 3 1 )/2                                                             } \\
 \noalign{\smallskip}
 1s2      &(1Se)& 0  & 2p&  1& -4.15050E+00 &  1.96 &  2~ P o       \\
 1s2      &(1Se)& 0  & 2p&  3& -4.14885E+00 &  1.96 &  2~ P o       \\
\hline
\end{tabular}
\end{table}

\begin{table}
\noindent{Table 3b. Sample table for calculated fine structure energy 
levels of C IV and Ni XXVI in $J\pi$ order. $Nlv$ is the total number 
of levels of the symmetry. \\
}
\begin{tabular}{rlrrlrl}
\hline
\noalign{\smallskip}
 & \multicolumn{3}{c}{Level} & \multicolumn{1}{c}{E(Ry)} & 
\multicolumn{1}{c}{$\nu$} & \multicolumn{1}{c}{$SL\pi$}  \\
\hline
            \noalign{\smallskip}
\multicolumn{7}{c}{C IV: $N_b$ = 98} \\
            \noalign{\smallskip}
 \multicolumn{7}{c}{Nlv=  9,~~~~~J pi =  1/2 e       } \\
 \noalign{\smallskip}
            \hline
            \noalign{\smallskip}
   1 & $1s2    (^1S^e_0)$ & $ 2s$ &  1/2 & -4.73899E+00 &  1.84 & $^2S e       $
 \\
   2 & $1s2    (^1S^e_0)$ & $ 3s$ &  1/2 & -1.98187E+00 &  2.84 & $^2S e       $
 \\
   3 & $1s2    (^1S^e_0)$ & $ 4s$ &  1/2 & -1.08239E+00 &  3.84 & $^2S e       $
 \\
   4 & $1s2    (^1S^e_0)$ & $ 5s$ &  1/2 & -6.81555E-01 &  4.84 & $^2S e       $
 \\
   5 & $1s2    (^1S^e_0)$ & $ 6s$ &  1/2 & -4.68304E-01 &  5.84 & $^2S e       $
 \\
\hline
\end{tabular}
\end{table}

\begin{table}
\noindent{Table 4. Calculated energies for the fine structure levels 
of $n$ = 2 and 3 complexes of Li-like ions. $N_b$ is the total number of 
levels obtained. The number in the first column specifies the energy 
order of the level in the $J\pi$ symmetry. \\
}
\begin{tabular}{rlrrll}
\hline
\noalign{\smallskip}
 & \multicolumn{2}{c}{Level}& \multicolumn{1}{c}{J} & \multicolumn{1}{c}{E(Ry)}
 & \multicolumn{1}{c}{$SL\pi$}  \\
\hline
            \noalign{\smallskip}
\multicolumn{6}{c}{C IV: $N_b$ = 97} \\
 \noalign{\smallskip}
            \hline
            \noalign{\smallskip}
   1 & $1s2    (^1S^e_0)$ & $ 2s$ &  1/2 & -4.73899E+00 &  $^2S e       $ \\
   2 & $1s2    (^1S^e_0)$ & $ 3s$ &  1/2 & -1.98187E+00 &  $^2S e       $ \\
   1 & $1s2    (^1S^e_0)$ & $ 2p$ &  1/2 & -4.15050E+00 &  $^2P o       $ \\
   2 & $1s2    (^1S^e_0)$ & $ 3p$ &  1/2 & -1.82526E+00 &   $^2P o       $ \\
   1 & $1s2    (^1S^e_0)$ & $ 3d$ &  3/2 & -1.78290E+00 &   $^2D e       $ \\
   1 & $1s2    (^1S^e_0)$ & $ 2p$ &  3/2 & -4.14885E+00 &   $^2P o       $ \\
\hline
\end{tabular}
\end{table}

\begin{table}
\noindent{Table 4 continues, \\
}
\begin{tabular}{rlrrll}
\hline
\noalign{\smallskip}
 & \multicolumn{2}{c}{Level}& \multicolumn{1}{c}{J} & \multicolumn{1}{c}{E(Ry)}
 & \multicolumn{1}{c}{$SL\pi$}  \\
\noalign{\smallskip}
 \hline
 \noalign{\smallskip}
 \multicolumn{6}{c}{ Mg X: $N_b$ = 98 } \\
 \noalign{\smallskip}
 \hline
 \noalign{\smallskip}
   1 & $1s2    (^1S^e_0)$ & $ 2s$ &  1/2 & -2.70076E+01 &  $^2S e       $ \\
   2 & $1s2    (^1S^e_0)$ & $ 3s$ &  1/2 & -1.16722E+01 &   $^2S e       $ \\
   1 & $1s2    (^1S^e_0)$ & $ 2p$ &  1/2 & -2.55546E+01 &   $^2P o       $ \\
   2 & $1s2    (^1S^e_0)$ & $ 3p$ &  1/2 & -1.12786E+01 &   $^2P o       $ \\
   1 & $1s2    (^1S^e_0)$ & $ 3d$ &  3/2 & -1.11259E+01 &   $^2D e       $ \\
   1 & $1s2    (^1S^e_0)$ & $ 2p$ &  3/2 & -2.55087E+01 &  $^2P o       $ \\
\hline
\end{tabular}
\end{table}

\begin{table}
\noindent{Table 4 continues, \\
}
\begin{tabular}{rlrrll}
\hline
\noalign{\smallskip}
 & \multicolumn{2}{c}{Level}& \multicolumn{1}{c}{J} & \multicolumn{1}{c}{E(Ry)}
 & \multicolumn{1}{c}{$SL\pi$}  \\
 \noalign{\smallskip}
 \hline
 \noalign{\smallskip}
 \multicolumn{6}{c}{Ti XX: $N_b$ = 98 } \\
 \noalign{\smallskip}
 \hline
 \noalign{\smallskip}
   1 & $1s2    (^1S^e_0)$ & $ 2s$ &  1/2 & -1.04794E+02 &  $^2S e       $ \\
   2 & $1s2    (^1S^e_0)$ & $ 3s$ &  1/2 & -4.58280E+01 &   $^2S e       $ \\
   1 & $1s2    (^1S^e_0)$ & $ 2p$ &  1/2 & -1.01885E+02 &   $^2P o       $ \\
   2 & $1s2    (^1S^e_0)$ & $ 3p$ &  1/2 & -4.50420E+01 &   $^2P o       $ \\
   1 & $1s2    (^1S^e_0)$ & $ 3d$ &  3/2 & -4.45516E+01 &   $^2D e       $ \\
   1 & $1s2    (^1S^e_0)$ & $ 2p$ &  3/2 & -1.01238E+02 &   $^2P o       $ \\
\hline
\end{tabular}
\end{table}

\clearpage

\begin{table}
\noindent{Table 5. $f$- and $A$-values for $n$ = 2 and 3 transitions in
Li-like ions. The calculated transition energies are replaced by the 
observed energies. The notation $a(b)$ means $a\times 10^b$. \\ }
\scriptsize
\begin{tabular}{llllrrrcl}
\noalign{\smallskip}
\hline
\noalign{\smallskip}
$C_i$ & $C_j$  & $T_i$ & $T_j$ & $g_i$ & $g_j$ & $E_{ij}$ &
$f$ & $A$ \\
 & & & & & & $(\AA)$ & & $(s^{-1})$ \\
 \noalign{\smallskip}
\hline
 \noalign{\smallskip}
\multicolumn{9}{c}{C IV} \\
\hline
 \noalign{\smallskip}
 $     2s     $ & $       2p     $ & $^2S^e$ & $^2P^o$ &  2    &  2    &   1550.8 &  9.501(-2) &  2.63(8) \\
 $     2s    $ & $        2p     $ & $^2S^e$ & $^2P^o$ &  2    &  4    &   1548.2 &  1.904(-1) &  2.65(8) \\
\multicolumn{2}{l}{$LS$} & $^2S^e$ & $^2P^o$ &  2 &  6 &  &  2.854(-1) &  2.64(8) \\
 & & & & & & & &  \\
 $     2s    $ & $        3p     $ & $^2S^e$ & $^2P^o$ &  2    &  2    &    312.5 &  6.810(-2) &  4.65(9) \\
\hline
\end{tabular}
\end{table}

\begin{table}
\noindent{Table 5. continues.  \\ }
\scriptsize
\begin{tabular}{llllrrrcl}
\noalign{\smallskip}
\hline
\noalign{\smallskip}
$C_i$ & $C_j$  & $T_i$ & $T_j$ & $g_i$ & $g_j$ & $E_{ij}$ &
$f$ & $A$ \\
 & & & & & & $(\AA)$ & & $(s^{-1})$ \\
 \noalign{\smallskip}
\hline
 \noalign{\smallskip}
\multicolumn{9}{c}{F VII} \\
\hline
 \noalign{\smallskip}
 $    2s     $ & $       2p      $ & $^2S^e$ & $^2P^o$ &  2    &  2    &    890.8 &  5.710(-2) &  4.80(8) \\
 $    2s     $ & $       2p      $ & $^2S^e$ & $^2P^o$ &  2    &  4    &    883.0 &  1.154(-1) &  4.93(8) \\
\multicolumn{2}{l}{$LS$} & $^2S^e$ & $^2P^o$ &  2 &  6 &  &  1.725(-1) &  4.89(8) \\
 & & & & & & & & \\
 $    2s      $ & $      3p      $ & $^2S^e$ & $^2P^o$ &  2    &  2    &    113.0 &  9.585(-2) &  5.01(10) \\
\hline
\end{tabular}
\end{table}

\begin{table}
\noindent{Table 5 continues. \\}
\scriptsize
\begin{tabular}{llllrrrcl}
\noalign{\smallskip}
\hline
\noalign{\smallskip}
$C_i$ & $C_j$  & $T_i$ & $T_j$ & $g_i$ & $g_j$ & $E_{ij}$ &
$f$ & $A$ \\
 & & & & & & $(\AA)$ & & $(s^{-1})$ \\
 \noalign{\smallskip}
\hline
 \noalign{\smallskip}
\multicolumn{9}{c}{Mg X} \\
 \noalign{\smallskip}
\hline
 \noalign{\smallskip}
 $2s   $ & $  2p   $ & $^2S^e$ & $^2P^o$ &  2    &  2    &    624.6 &  4.078(-2) &  6.97(8) \\
 $2s   $ & $  2p   $ & $^2S^e$ & $^2P^o$ &  2    &  4    &    609.5 &  8.384(-2) &  7.53(8) \\
 \multicolumn{2}{l}{$LS$} & $^2S^e$ & $^2P^o$ &  2 &  6 &  &  1.246(-1) &  7.34(8) \\
 & & & & & & & & \\
 $2s   $ & $  3p   $ & $^2S^e$ & $^2P^o$ &  2    &  2    &     57.9 &  1.094(-1) &  2.18(11) \\
\hline
\end{tabular}
\end{table}

\begin{table}
\noindent{Table 5 continues. \\}
\scriptsize
\begin{tabular}{llllrrrcl}
\noalign{\smallskip}
\hline
\noalign{\smallskip}
$C_i$ & $C_j$  & $T_i$ & $T_j$ & $g_i$ & $g_j$ & $E_{ij}$ &
$f$ & $A$ \\
 & & & & & & $(\AA)$ & & $(s^{-1})$ \\
 \noalign{\smallskip}
\hline
 \noalign{\smallskip}
\multicolumn{9}{c}{S XIV} \\
\hline
 \noalign{\smallskip}
 $ 2s      $ & $   2p      $ & $^2S^e$ & $^2P^o$ &  2    &  2    &    445.8 &  2.951(-2) &  9.90(8) \\
 $ 2s      $ & $   2p      $ & $^2S^e$ & $^2P^o$ &  2    &  4    &    417.6 &  6.335(-2) &  1.21(9) \\
 \multicolumn{2}{l}{$LS$} & $^2S^e$ & $^2P^o$ &  2 &  6 &  &  9.285(-2) &  1.13(9) \\
 & & & & & & & & \\
 $ 2s      $ & $   3p      $ & $^2S^e$ & $^2P^o$ &  2    &  2    &     30.5 &  1.198(-1) &  8.61(11) \\
\hline
\end{tabular}
\end{table}

\begin{table}
\noindent{Table 5 continues. \\}
\scriptsize
\begin{tabular}{llllrrrcl}
\noalign{\smallskip}
\hline
\noalign{\smallskip}
$C_i$ & $C_j$  & $T_i$ & $T_j$ & $g_i$ & $g_j$ & $E_{ij}$ &
$f$ & $A$ \\
 & & & & & & $(\AA)$ & & $(s^{-1})$ \\
 \noalign{\smallskip}
\hline
 \noalign{\smallskip}
\multicolumn{9}{c}{Ti XX} \\
\hline
 \noalign{\smallskip}
 $ 2s      $ & $   2p      $ & $^2S^e$ & $^2P^o$ &  2    &  2    &    308.9 &  2.093(-2) &  1.46(9) \\
 $ 2s      $ & $   2p      $ & $^2S^e$ & $^2P^o$ &  2    &  4    &    259.6 &  5.040(-2) &  2.49(9) \\
 \multicolumn{2}{l}{$LS$} & $^2S^e$ & $^2P^o$ &  2 &  6 &  &  7.130(-2) &  2.11(9) \\
 & & & & & & & & \\
 $ 2s      $ & $   3p      $ & $^2S^e$ & $^2P^o$ &  2    &  2    &     15.3 &  1.281(-1) &  3.67(12) \\
\hline
\end{tabular}
\end{table}

\clearpage

\begin{table}
\noindent{Table 6. Comparison of BPRM A-values with those in the NIST 
compilation, Refs: a-Johnson et al (1996), b-Yan et al. (1998). \\}
\scriptsize
\begin{tabular}{llllrrrll}
\noalign{\smallskip}
\hline
\noalign{\smallskip}
$C_i$ & $C_j$  & $T_i$ & $T_j$ & $g_i$ & $g_j$ & $E_{ij}$ &
$A(s^{-1})$ & $A(s^{-1})$ \\
& & & & & & (Ry) & present & NIST \\
 \noalign{\smallskip}
\hline
\noalign{\smallskip}
\multicolumn{9}{c}{C IV} \\
 \noalign{\smallskip}
\hline
 \noalign{\smallskip}
 $  2s     $ & $    2p     $ & $^2S^e$ & $^2P^o$ &  2    &  2    &   1550.8 &  2.63(8) &  2.64(8) \\
 $  2s     $ & $    2p     $ & $^2S^e$ & $^2P^o$ &  2    &  4    &   1548.2 &  2.65(8) &  2.65(8) \\
 & & & & & & & & \\
 $  2s     $ & $    3p     $ & $^2S^e$ & $^2P^o$ &  2    &  2    &    312.5 &  4.65(9) &  4.63(9) \\
 $  2s     $ & $    3p     $ & $^2S^e$ & $^2P^o$ &  2    &  4    &    312.4 &  4.64(9) &  4.63(9) \\
\hline
\end{tabular}
\end{table}

\begin{table}
\noindent{Table 6 continues. \\}
\scriptsize
\begin{tabular}{llllrrrcc}
\noalign{\smallskip}
\hline
\noalign{\smallskip}
$C_i$ & $C_j$  & $T_i$ & $T_j$ & $g_i$ & $g_j$ & $E_{ij}$ &
$A(s^{-1})$ & $A(s^{-1})$ \\
& & & & & & (Ry) & present & NIST \\
 \noalign{\smallskip}
\hline
 \noalign{\smallskip}
\multicolumn{9}{c}{Si XII} \\
 \noalign{\smallskip}
\hline
 \noalign{\smallskip}
 $ 2s      $ & $   2p      $ & $^2S^e$ & $^2P^o$ &  2    &  2    &    520.4 &  8.44(8) &  8.59(8) \\
 $ 2s      $ & $   2p      $ & $^2S^e$ & $^2P^o$ &  2    &  4    &    499.3 &  9.59(8) &  9.56(8) \\
 & & & & & & & & \\
 $ 2p      $ & $   3s      $ & $^2P^o$ & $^2S^e$ &  2    &  2    &     45.5 &  6.73(10) &  6.68(10) \\
 $ 2p      $ & $   3s      $ & $^2P^o$ & $^2S^e$ &  4    &  2    &     45.7 &  1.37(11) &  1.32(11) \\
\hline
\end{tabular}
\end{table}

\clearpage

\begin{table*}
\noindent{Table 7. Sample set of $f$- and $A$-values of O VI in $J\pi$
order as obtained from BPRM calculations.  \\ }
\scriptsize
\begin{tabular}{rrlllll}
\noalign{\smallskip}
\hline
\noalign{\smallskip}
\multicolumn{2}{c}{  8} & \multicolumn{5}{l}{  3} \\
\multicolumn{7}{l}{~~~~ 1~~~~ 0~~~~~~~ 1~~~~ 1} \\
   9 &    9 & $E_i(Ry)$ & $E_j(Ry)$ & $gf_L$ & $ S  $ & $A_{ji}(sec^{-1})$ \\
\noalign{\smallskip}
   1&    1& -1.01495E+01& -9.27099E+00& -1.318E-01&  4.501E-01&  4.085E+08 \\
   1&    2& -1.01495E+01& -4.08326E+00& -1.778E-01&  8.793E-02&  2.628E+10 \\
   1&    3& -1.01495E+01& -2.28406E+00& -4.934E-02&  1.882E-02&  1.226E+10 \\
   1&    4& -1.01495E+01& -1.45741E+00& -2.104E-02&  7.262E-03&  6.385E+09 \\
   1&    5& -1.01495E+01& -1.01008E+00& -1.111E-02&  3.647E-03&  3.726E+09 \\
\hline
\end{tabular}
\end{table*}

\end{document}